\documentclass{sig-alt-full}
\usepackage{algorithmic}

\def\more-auths{%
\end{tabular}
\begin{tabular}{c}}

\begin{document}
\conferenceinfo{SPAA'04,} {June 27--30, 2004, Barcelona, Spain.} 
\CopyrightYear{2004} 
\crdata{1-58113-840-7/04/0006}
\title{Bi-criteria Algorithm for Scheduling Jobs on Cluster Platforms \titlenote{Authors are members of the APACHE project supported by CNRS, INPG, INRIA, UJF}} 
%
%

\numberofauthors{2}
%

\author{
%
\alignauthor Pierre-Fran{\c c}ois Dutot\\
       \affaddr{ID-IMAG}\\
       \affaddr{51 avenue Jean Kuntzmann}\\
       \affaddr{38330 Montbonnot Saint-Martin, France}\\
       \email{pfdutot@imag.fr}\\
\alignauthor Lionel Eyraud\\
       \affaddr{ID-IMAG}\\
       \affaddr{51 avenue Jean Kuntzmann}\\
       \affaddr{38330 Montbonnot Saint-Martin, France}\\
       \email{Lionel.Eyraud@imag.fr}\\
%
%
\more-auths
\alignauthor Gr{\' e}gory Mouni{\' e}\\
       \affaddr{ID-IMAG}\\
       \affaddr{51 avenue Jean Kuntzmann}\\
       \affaddr{38330 Montbonnot Saint-Martin, France}\\
       \email{Gregory.Mounie@imag.fr}
\alignauthor Denis Trystram\\
       \affaddr{ID-IMAG}\\
       \affaddr{51 avenue Jean Kuntzmann}\\
       \affaddr{38330 Montbonnot Saint-Martin, France}\\
       \email{Denis.Trystram@imag.fr}
}
\date{27 June 2004}

\maketitle

\begin{abstract}
  We describe in this paper a new method for building an efficient algorithm
  for scheduling jobs in a cluster. Jobs are considered as parallel
  tasks (PT) which can be scheduled on any number of processors. The
  main feature is to consider two criteria that are optimized
  together.  These criteria are the {\it makespan} and the weighted
  minimal average completion time ({\it minsum}). They are chosen for
  their complementarity, to be able to represent both user-oriented
  objectives and system administrator objectives.
  
  We propose an algorithm based on a batch policy with increasing batch
  sizes, with a smart selection of jobs in each batch. This algorithm
  is assessed by intensive simulation results, compared to a new lower
  bound (obtained by a relaxation of ILP) of the optimal schedules for
  both criteria separately. It
  is currently implemented in an actual real-size cluster platform.
\end{abstract}

\category{F.2.2}{Analysis of Algorithms and Problem Complexity}{Nonnumerical Algorithms and Problems}[Sequencing and scheduling]
\category{D.4.1}{Operating Systems}{Process management}[Scheduling, Concurrency]

\terms{Algorithms, Management}

\keywords{Parallel Computing, Algorithms, Scheduling, Parallel Tasks, Moldable Tasks, Bi-criteria}

\section{Introduction} 

\subsection{Cluster computing}

The last few years have been characterized by huge technological 
changes in the area of parallel and distributed computing.
Today, powerful machines are available at low price everywhere in the world. 
The main visible line of such changes is the large spreading
of {\it clusters} which consist in a collection of tens or hundreds 
of standard almost identical processors connected
together by a high speed interconnection network \cite{CullerSingh99}.
The next natural step is the extension to local sets of
clusters or to geographically distant grids \cite{FosterKesselman}.

In the last issue of the Top500 ranking (from November 2003
\cite{top500}), 52 networks of workstations (NOW) of different kinds
were listed and 123 entries are clusters sold either by IBM, HP or
Dell. Looking at previous rankings we can see that this number (within
the Top500) approximately doubled each year.

This democratization of clusters calls for new practical
administration tools.  Even if more and more applications are running
on such systems, there is no consensus towards an universal way of
managing efficiently the computing resources.  Current available
scheduling algorithms were mainly created to provide schedules with
performance guaranties for the makespan criterion (maximum execution
time of the last job), however most of them are pseudo-polynomial,
therefore the time needed to run these algorithms on real instances
and the difficulty of their implementation is a drawback for a more
popular use.

  We present in this paper a new method for scheduling the jobs
submitted to a cluster inspired by several existing theoretically well-founded
algorithms.  This method has been assessed on simulations and it is
currently tested on actual conditions of use on a large cluster
composed by 104 bi-processor machines from Compaq (this cluster --
called Icluster2 -- was ranked 151 in the Top500 in June 2003).

To achieve reasonable performance within reasonable time, we decided
to build a fast algorithm which has the best features of existing
ones. However, to speed up the algorithm a guaranteed performance
ratio cannot be achieved, thus we concentrate on the average ratio on
a large set of generated test instances. These instances are representative of
jobs submitted on the Icluster \cite{DRTipdps04}.

\subsection{Related approaches}

Some scheduling algorithms have been developed for classical
parallel and distributed systems of the last generations.  Clusters
introduce new characteristics that are not really taken into account
into existing scheduling modules, namely, unbalance between
communications and computations -- communications are relatively large
-- or on-line submissions of jobs.

Let us present briefly some schedulers used in actual systems: the
basic idea in job schedulers \cite{henderson95:_job} is to queue jobs
and to schedule them one after the other using some simple rules like
FCFS (First Come First Served) with priorities. MAUI scheduler
\cite{jackson01:_core_algor_maui_sched} extends the model with
additional features like fairness and backfilling. 

 AppleS is an application level scheduler system for grid. It is used
to schedule, for example, an application composed of a large set of
independent jobs with shared data input files
\cite{casanova00:_apples_param_sweep_templ}. It selects resources
efficiently and takes into account data distribution time. It is
designed for grid environment.

There exist other parallel environments with a more general spectrum
(heterogeneous and versatile execution platform) like Condor
\cite{Livny} or with special capabilities like processus migration,
requiring system-level implementation like Mosix \cite{Barak}.
However, in these environments scheduling algorithms are online
algorithms with simple rules.

\subsection{Our approach}

As no fast and flexible scheduling systems are available today for
clusters, we started two years ago to develop a new system based on a
sound theoretical background and a significant practical experience
of managing a big cluster (Icluster1, a 225 PC machine arrived in 2001
in our lab).  It is based on the model of {\it parallel tasks}
\cite{Feitelsonsurvey} which are independent jobs submitted by the
users.


We are interested here in optimizing simultaneously two criteria,
namely the minsum ($\Sigma C_i$) which is usually targeted by the users
who all want to finish their jobs as soon as possible, and the makespan
($C_{max}$) which is rather a system administrator objective
representing the total occupation time of the platform.

 There exist algorithms for each
criterion separately; we propose here a bi-criteria algorithm to
optimize the $C_{max}$ and $\Sigma C_i$ criteria simultaneously. The
best existing algorithm for minimizing the makespan off-line (all jobs
are available at the beginning) has a $3/2 + \epsilon$ guaranty
\cite{chapter28DMT}. We can derive easily an on-line batch version by
using the general framework of \cite{SWW95} leading to an
approximation ratio of $3 + \epsilon$. For the other criterion, the
best result is $8$ for the unweighted case and $8.53$ for the weighted
case \cite{SchwiegelshohnEtAl98}.  Using a nice generic framework
introduced by Hall et al.\cite{HaScSW97}, a ($12$;$12$) approximation
can be obtained at the cost of a big complexity which impedes the use
of such algorithms.


The paper is organized as follows: In the next section, we will
introduce the definitions and models used in all the paper.
The
algorithm itself is described in section \ref{sec:algo}, along with
the lower bound which is used in the experiments.
 The experimental setting and the results
are discussed in section \ref{sec:experiences}.
 Finally we
will conclude in section \ref{sec:conclu} with a discussion on on-going
works.


\section{Context and Definition} 
\label{sec:context}

\subsection{Architectural and Computing Models}

The target execution support that we consider here is a cluster
composed by a collection of a medium number of SMP or simple PC
machines (typically several dozens or several hundreds of nodes). 
The nodes are fully connected and homogeneous.

\begin{figure}[ht]
\begin{center}
\includegraphics[scale=0.5]{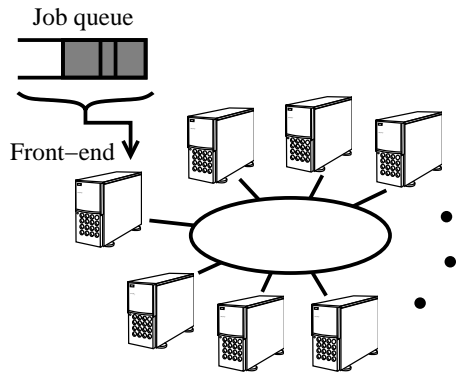}
\end{center}
\caption{Job submission in clusters.} \label{lightgrid}
\end{figure}

The submissions of jobs is done by some specific nodes by the way of
several priority queues as depicted in Figure \ref{lightgrid}.  No other submission is allowed.


Informally, a Parallel Task (PT) is a {\it task} that gathers
elementary operations, typically a numerical routine or a nested loop,
which contains itself enough parallelism to be executed by more than
one processor.  We studied scheduling of one specific kind of PT,
denoted as {\em moldable jobs} according to the classification of Feitelson et al.
\cite{Feitelson99}. The number of processors to
execute a moldable job is not fixed but determined before the
execution, as opposed to {\em rigid jobs} where the number of
processors is fixed by the user at submission time.  In any case, the
number of processors does not change until the completion of the job.

For historical reasons, most of submitted jobs are rigid.  However,
intrinsically, most parallel applications are moldable. An application
developer does not know in advance the exact number of processors
which will be used at run time. Moreover, this number may vary with
the input problem size or number of available nodes.  This is also
true for many numerical parallel libraries.  The main exception to
this rule is when a minimum number of processors is required because
of time, memory or storage constraints.

The main restriction in a systematic use of the moldable character is
the need for a practical and reliable way to estimate (at least
roughly) the parallel execution time as function of the number of
processors.  Most of the time, the user has this knowledge but does
not provide it to the scheduler, as it is not taken into account by
rigid jobs schedulers. This is an inertia factor against the more
systematic use of such models, as the users habits have to be changed.

Our algorithm proposes, thanks to moldability, to efficiently decrease
average response time (at the users request) while keeping computing
overhead and idle time as low as possible (at the system
administrators request).

\subsection{Scheduling on clusters} 
\label{sec:etatdelart}


The main objective function used historically is the {\it makespan}.
This function measures the ending time of the schedule,
i.e., the latest completion time over all the tasks. However, this
criterion is valid only if we consider the tasks altogether and
from the viewpoint of a single user. If the tasks have been
submitted by several users, other criteria can be considered. Let us
present briefly the two criteria:
  
\begin{itemize}
\item
Minimization of the {\it makespan}
($C_{max} = max(C_j)$ where the completion time $C_j$ is equal to $\sigma(j) + p_j(nbproc(j))$).
$p_j$ represents the execution time of task $j$,
$\sigma$ function is the starting time and $nbproc$ function is the processor number
(it can be a vector in the case of specific allocations for heterogeneous processors).
\item
Minimization of the average completion time ($\Sigma C_i$)
\cite{ShaTur99,AEBFJK00} and its variant weighted completion time ($\Sigma \omega_i C_i$).
Such a weight may allow us to distinguish some tasks from each other (priority
for the smallest ones, etc.).
\end{itemize}

In a production cluster context, the jobs are submitted at any
time. Models were the characteristics of the tasks (duration, release
date, etc) are only known when the task is submitted are called {\it
  on-line} as opposed to the {\it off-line} models were all the tasks
are known and available at all times.
It is possible to schedule jobs on-line with a constant competitive
ratio for $C_{max}$. The idea is to schedule jobs by batches depending
on their arrival time. An arriving job is scheduled in the next
starting batch.  This simple rule allows constant competitive ratio in
the on-line case if a single batch may be scheduled with a constant
competitive ratio $\rho$. 

Roughly, the last batch starts after the
last task arrival date. By definition, all the tasks scheduled in a
batch are scheduled in less than $\rho C_{max}^*$, where $C_{max}^*$
is the optimal off-line makespan of the complete instance. The length
of the previous last batch is then lower than $\rho
C_{max}^*$. Moreover, the length of the last batch, plus the starting
time of the previous last batch (at which none of the tasks of the
last batch were released) is less than $\rho$ times the length of the optimal
on-line makespan. 

As the on-line makespan is larger than the off-line
makespan, the total schedule length is less than $2\rho$ times the
on-line optimal makespan.
This is how the off-line $3/2+\epsilon$ algorithm
is turned into an on-line $3+\epsilon$ algorithm as we said in the
introduction.




\section{A new bicriteria efficient solution}
\label{sec:algo}

\subsection{Rationale} 

Studying some extreme instances and their optimal schedules for the
minsum criterion, gave us an insight on the shape of the schedules we
had to build. For example, if all the tasks are perfectly moldable
(when the work does not depend on the number of processors) the
optimal solution is to schedule all the tasks on all processors in
order of increasing area. This example shows that the minsum criterion
tends to give more importance to the smaller tasks.

Previous algorithms presented in the literature are also designed to
take into account this global structure of scheduling the smaller
tasks first. Shmoys et al. \cite{HaScSW97} used a batch scheduling with
batches of increasing sizes. The batch length is doubled at each step,
therefore only the smaller tasks are scheduled in the first batches.

Existing makespan algorithms for moldable tasks are also designed with
a common structure of shelves (were all tasks start at the same time)
which is a relaxed version of batches. See for example
\cite{MounieRapineTrystram1999} or
\cite{chapter28DMT} for schedules with 2 shelves.

\begin{figure}[ht]
\begin{center}
\includegraphics[width=0.9\linewidth]{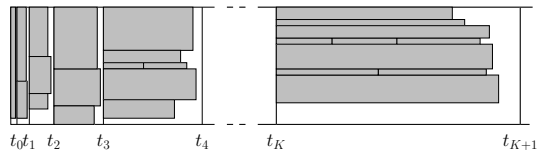}
\end{center}
\caption{Principle of the algorithm.} \label{fig:algo-principle}
\end{figure}

Our algorithm was built with this structure in mind: stacking tasks in
shelves of increasing sizes with the additional possibility of
shuffling these shelves if necessary.  However, our main motivation
was to design a fast algorithm for the management of some clusters of
a big regional grid in Grenoble. Our algorithm does not have a known
performance guaranty on the worst cases, however we tested its
behavior on a set of generated instances which simulate real jobs
submitted on our local clusters. The principle of the algorithm is
shown in Figure~\ref{fig:algo-principle}.

\subsection{Algorithm}

More formally, we detail below the algorithm starting with the
input describing the instances:
\begin{itemize}
\item $n$ tasks available at time $0$
\item $p_i(k)$ the processing time of task $i$ on $k$
processors
\item $w_i$ is its weight
\item $m$ the number of processors
\end{itemize}

\noindent
\fbox{%
\begin{minipage}{232pt}
\begin{algorithmic}
\STATE Compute the approximate $C_{max}^*$ with the dual approximation
algorithm.
\STATE $t_{min} = \min_{i,j} \{p_i(j)\}$
\STATE $K= \lfloor \log_2 \left ( \frac{C_{max}^*}{t_{min}} \right )
\rfloor$
\FOR {$j = 0..K+1$}
\STATE $t_j = \frac{C_{max}^*}{2^{K-j}}$
\ENDFOR
\STATE $T = \{1..n\}$
\FOR {$j=0..K$}
\STATE $S=\{i \in T\mbox{ such that }\exists j,\ p_i(j) \leq t_{j} \}$
\STATE Merge the small sequential tasks sorted by decreasing weight.
\STATE Select the set $S_j \subseteq S$ of tasks to schedule in the
current batch (using a knapsack).
\STATE Schedule the batch between $t_j$ and $t_{j+1}$.
\STATE Remove $S_j$ from $T$.
\ENDFOR
\STATE Compact the schedule with a list algorithm using the batch ordering.
\end{algorithmic}
\end{minipage}
}

\bigskip

First, our algorithm calls a dual approximation makespan algorithm
(defined in \cite{chapter28DMT}) to determine an
approximation of the optimal makespan of the instance. With this value
$C_{max}^*$ and the smallest possible duration of a task $t_{min}$, we
compute the smallest useful batch size $t_0$ (such that at least one
task can be done) and $K+1$ the number of batches. The values $t_j$
are the length of our batches. For every $j$, $t_{j+1}$ is twice the
value of $t_j$.

The main loop of the algorithm corresponds to the selection of the jobs to be
scheduled in the current batch. We first select the tasks which are
not too long to run in the batch. If there are several tasks that can
be run in less than half the batch size on one processor, we can merge
some of these tasks by stacking them together. In order to have as
much weight as possible, this merge is done by decreasing weight order. 

The next step is to run a knapsack selection, written with integer
dynamic programming. We want to maximize the sum of the weight of the
selected tasks while using at most $m$ processors. The allocation of
the task $i$ is $allot_i$, the smallest allocation that fits (in length) into the
batch. Values of $W(i,j)$ are initialized to $-\infty$ for $j<0$ and
$0$ otherwise.
For $i$ going from $1$ to $n$ and for $j$ going from $1$ to
$m$, we compute:
$$
W(i,j)=max \left ( W(i-1,j) , W(i-1,j-allot_i) + w_i \right )
$$
The largest $W(n,\cdot)$ is the maximum weight that can be done in the batch.
The complexity of this knapsack is $O(mn)$. 

The first schedule is simple: we start all the selected tasks of one batch
at the same time. A straightforward improvement is to start a task
at an earlier time if all the processors it uses are idle. A further
improvement is to use a list algorithm with the batch ordering
and a local ordering within the batches, as it allows to change the
set of processors alloted to the tasks.

Finally, an additional optimization step is used. The batch order is
shuffled several times and the best resulting compact schedule is kept.
This only leads to small improvements. 

The overall complexity of this algorithm is $O(mnK)$.


\subsection{Lower Bound}
\label{sec:lowerbound}


In order to assess this algorithm with experiments, for each instance
we need to know the value of an optimal solution. But since the
problem is NP-Hard in the strong sense, computing an optimal solution
in reasonable time is impossible. We are thus looking for good lower
bounds. 

For $C_{max}$ a good lower bound may easily be obtained by dual
approximation \cite{chapter28DMT}. For $\Sigma C_i$ the lower bound is
computed by a relaxation of a Linear Programming formulation of the
problem. This formulation is not intended to yield a feasible
schedule, but rather to express constraints that are necessarily
respected by every feasible schedule.
For this formulation, we divided the time horizon into 
several intervals $I_j = (t_j, t_{j+1}]$ with $0\leq j \leq K$.
The  values of the $t_j$ and the value of $K$ are defined as in the
previous section.

Once the time division is fixed, we consider the decision variables
$x_{i,j}=1$ if
and only if task $i$ ends within $I_j$ (i.e. between $t_j$ and
$t_{j+1}$), and $x_{i,j}=0$ otherwise.

For each task $i$ and each interval $j$, we can also compute the
minimal area occupied by task $i$ if it ends before $t_{j+1}$:
$$S_{i,j} = \min_{1\leq k \leq m} \{ k p_i(k) \mbox{ such that } p_i(k) \leq t_{j+1}
\}$$
If the set is empty, let $S_{i,j} = +\infty$.

With these values, we can give the formulation of the problem:

$$\begin{array}{ll}
\text{Minimize} & \sum_{i,j} w_i t_j x_{i,j}\\
\text{Subject to} & \forall i, \quad \sum_j x_{i,j} \geq 1\\
                  & \forall j, \quad \sum_{0\leq l\leq j} \sum_i
                  S_{i,l}x_{i,l} \leq m t_{j+1}\\
                  & \forall i, \forall j, \quad x_{i,j} \in \{0, 1\}
\end{array}$$
   
The first constraint expresses that every task should be performed at
least once. The minimization criterion implies that no task will be
performed more than once: if $x_{i,j}$ and $x_{i, j'}$ are equal to
one, we get a better, yet still feasible solution by setting one of
them to zero.

The second constraint is a surface argument. For each interval $I_j$, we
consider the tasks that end before or in this interval (they end in
$I_l$, for $l \leq j$). By definition, a task $i$ ending in interval
$l$ takes up a surface at least $S_{i,l}$. The sum of all these
surfaces has to be smaller than the total surface between time $0$
and time $t_{j+1}$, which is $m t_{j+1}$. This is obviously
optimistic, because it does not take into account collisions between
tasks: scheduling according to this formulation might require
more than $m$ processors.

Both of these constraints are satisfied by every feasible
schedules, so for every feasible schedule $S$, there is a solution $R$
to this linear program. Since for each job $i$, $\sum_{j}t_j x_{i,j}
\leq C_i$, the objective function of $R$ is not larger than the minsum
criterion of the schedule $S$. In particular, every optimal schedule
yields a solution to the linear program, so the optimal value of the
objective function is always smaller than the optimal value of the
minsum criterion of the scheduling problem. This still holds when
considering the relaxed problem, where $x_{i,j}$ is in $[0; 1]$. The
lower bound might be weaker, but is much faster to compute.


\section{Experiments} 
\label{sec:experiences}

\subsection{Experimental setting}

The experimental simulations presented here were performed with an
ad-hoc program. Each experience is obtained by 40 runs; for each run
tasks are generated in an off-line manner, then given as an input to
the scheduling algorithm and to the linear solver which computes a
lower bound for this instance. Comparison between the two results
yields a performance ratio, and the average ratio for the whole set of
runs is the result of the experiments.

The runs were made assuming a cluster of 200 processors, and a
number of tasks varying from 25 to 400. In order to describe a
mono-processor task, only its computing time is needed. A moldable
task is described by a vector of $m$ processing times (one per number
of processor alloted to the task). We used
two different models to generate the tasks. The first one generates
the sequential processing times of the tasks, and the second one uses
a parallelism model to derive all the other values.

Two different sequential workload type were used: uniform and mixed
cases. For all uniform cases, sequential times were generated
according to an uniform distribution, varying from $1$ to $10$. For
mixed cases, we introduce two classes : small and large tasks. The
random values are taken with gaussian distributions centered
respectively on 1 and 10, with respective standard deviations of 0.5
and 5, the ratio of small tasks being 70\%.

Modeling the parallelism of the jobs was done in two different ways.
In the first, successive processing times were computed with the
formula $ p_i(j) = p_i(j-1) \frac{X + j}{1 + j} $, where $X$ is a
random variable between $0$ and $1$. Depending on the distribution of
$X$, tasks generated are highly parallel (with a quasi-linear speedup)
or weakly parallel (with a speedup close to $1$). Respectively highly
and weakly parallel are generated using gaussian distribution centered
on 0.9, and 0.1, and with a standard deviation of 0.2. Any random value
smaller than $0$ and larger than $1$ are ignored and recomputed.
According to the usual parallel program behavior, this method
generates monotonic tasks, which have decreasing execution times and
increasing work with $k$. For the mixed cases, the small tasks are
weakly parallel and the large tasks are highly parallel.

The second way of modeling parallelism was done according to a model
from Cirne and Berman~\cite{CirneBerman}, which relies on a survey
about the behavior of the users in a computing center. Only the
$uniform(1,10)$ sequential time model is used for theses tasks.

To evaluate our algorithm, we use the lower bound (cf section
\ref{sec:lowerbound}) as reference. Some simple "standard" algorithms are
used to compare the behavior and efficiency of our approach.

\begin{description}
\item[Gang :] Each task is scheduled on all processors. The tasks are
  sorted using the ratio of the weight over the execution time. This
  algorithm is optimal for instances with linear speedup.
\item[Sequential :] Each tasks is scheduled on a single processor. A
  list algorithm is used, scheduling large processing time first
  (LPTF).
\item[List Graham:] All the 3 algorithms are multiprocessor list
  scheduling \cite{GarGra75}. Every tasks is alloted using the number
  of processor selected by \cite{chapter28DMT}. This should lead to a
  very good average performance ratio with respect to the $C_{max}$
  criterion. Only the order of the list is changing between the three
  algorithms :
  \begin{itemize}
  \item the first one keep the order of \cite{chapter28DMT}, listing
    first task of the large shelf then the tasks of small shelf then
    the small tasks,
  \item weighted largest processing time first (LPTF), a classical
    variant, with a very god behavior for $C_{max}$ criterion, but the
    tasks are in fact sorted using the ratio between weighted and
    their execution time.
  \item smallest area first (SAF), almost the opposite of LPTF, the
    tasks are sorted according to their area (number of processors
    $\times$ execution time). The goal is to improve the average
    performance ratio for the $\sum w_i C_i$ criterion.
  \end{itemize}
\end{description}

In all experiments, task priority is a random value taken from an
uniform distribution between 1 and 10.

\subsection{Simulation results}


The results of the simulation runs are given in all the following
figures, plotting the minimum, maximum and average values for
$C_{max}$ and $\sum w_i C_i$. The average of the competitive ratio is
computed by dividing the sum of the execution times over the sum of
the lower bounds for every point \cite{Jain_chapRatioGame}.  Every
workload type are represented separately. The same scale is
represented for identical criterion between the workload type.

\begin{figure}[htbp]
\begin{center}
  \includegraphics[scale=0.70]{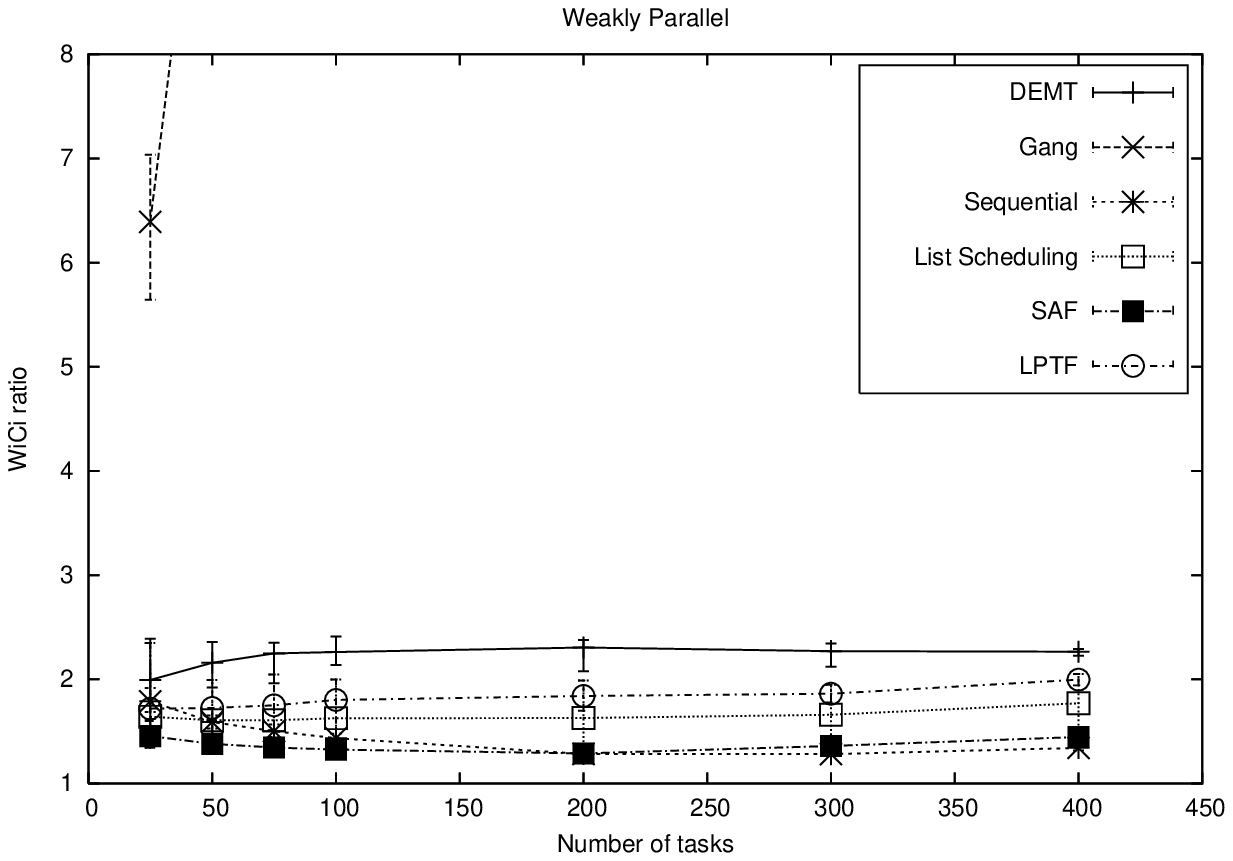}
  \includegraphics[scale=0.70]{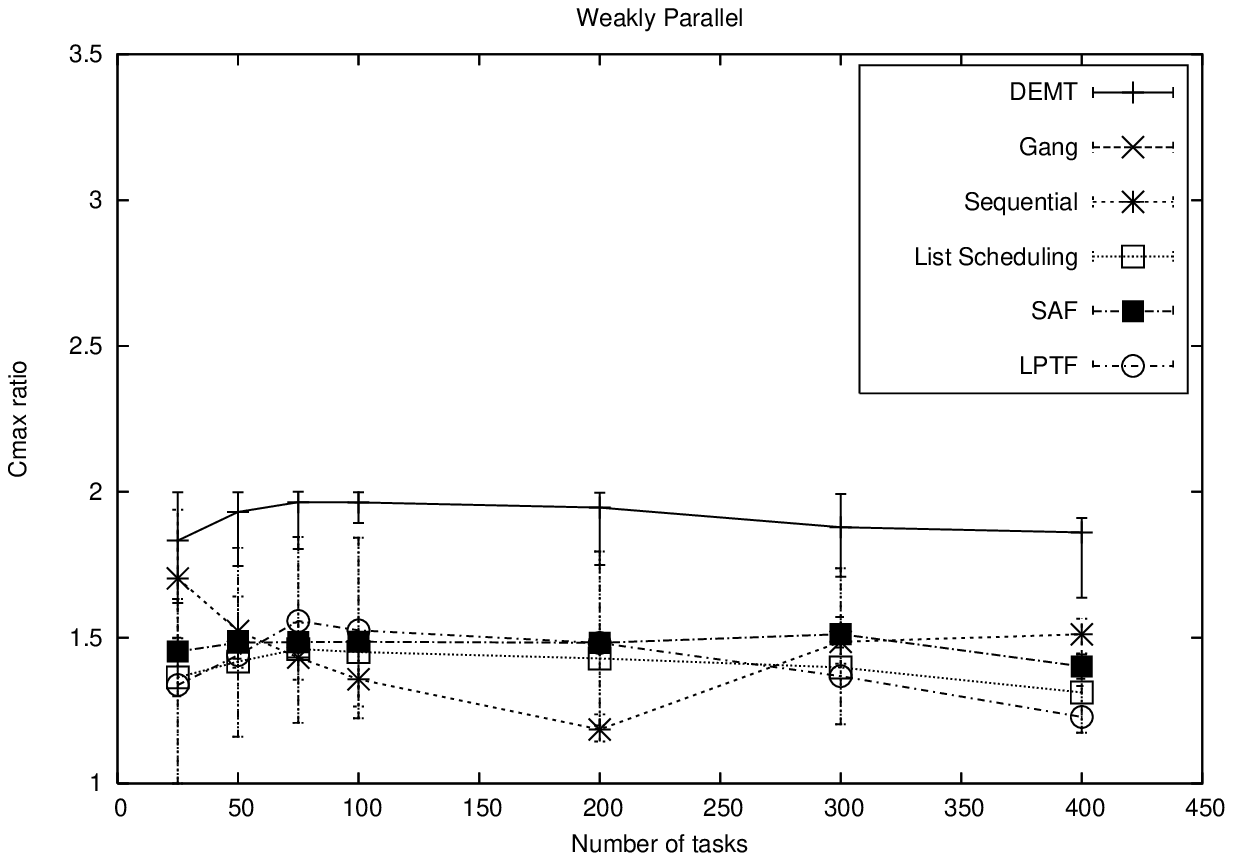}
\end{center}
\caption{Performance ratio for the simulation on 200 processors, weakly parallel tasks}
\label{fig:courbe_weak}
\end{figure}

The tasks of Figure \ref{fig:courbe_weak} are weakly parallel.
This is the worst case for our algorithm as it spends resources to
accelerate completion of small and high priority parallel tasks. These
resources are thus spend without much gain. Note that Gang scheduling
does not appear in the presented range for $C_{max}$, as Gang always
has a very big ratio in this case.

As expected, the average performance ratio for our algorithm is worse
than all other algorithms except Gang. Nevertheless, the performance
ratio for $C_{max}$ is no more than $2$. All other algorithms
have an average performance ratio around $1.5$. The difference is
large enough to influence also the results for the minsum
criterion. From this case we may deduce that for most cases, our
algorithms will not be much worse than a performance ratio of $2$ for
both criterion.

\begin{figure}[p]
\begin{center}
  \includegraphics[scale=0.70]{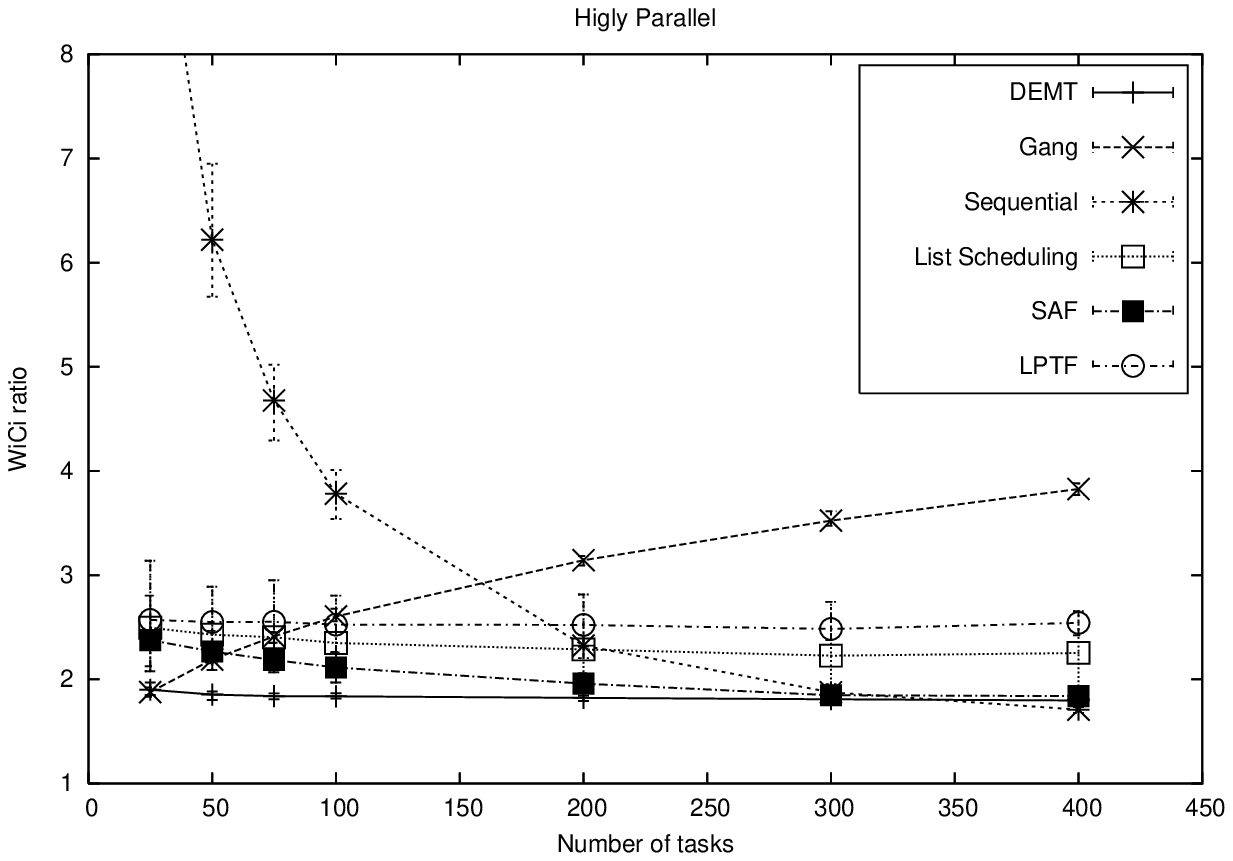}
  \includegraphics[scale=0.70]{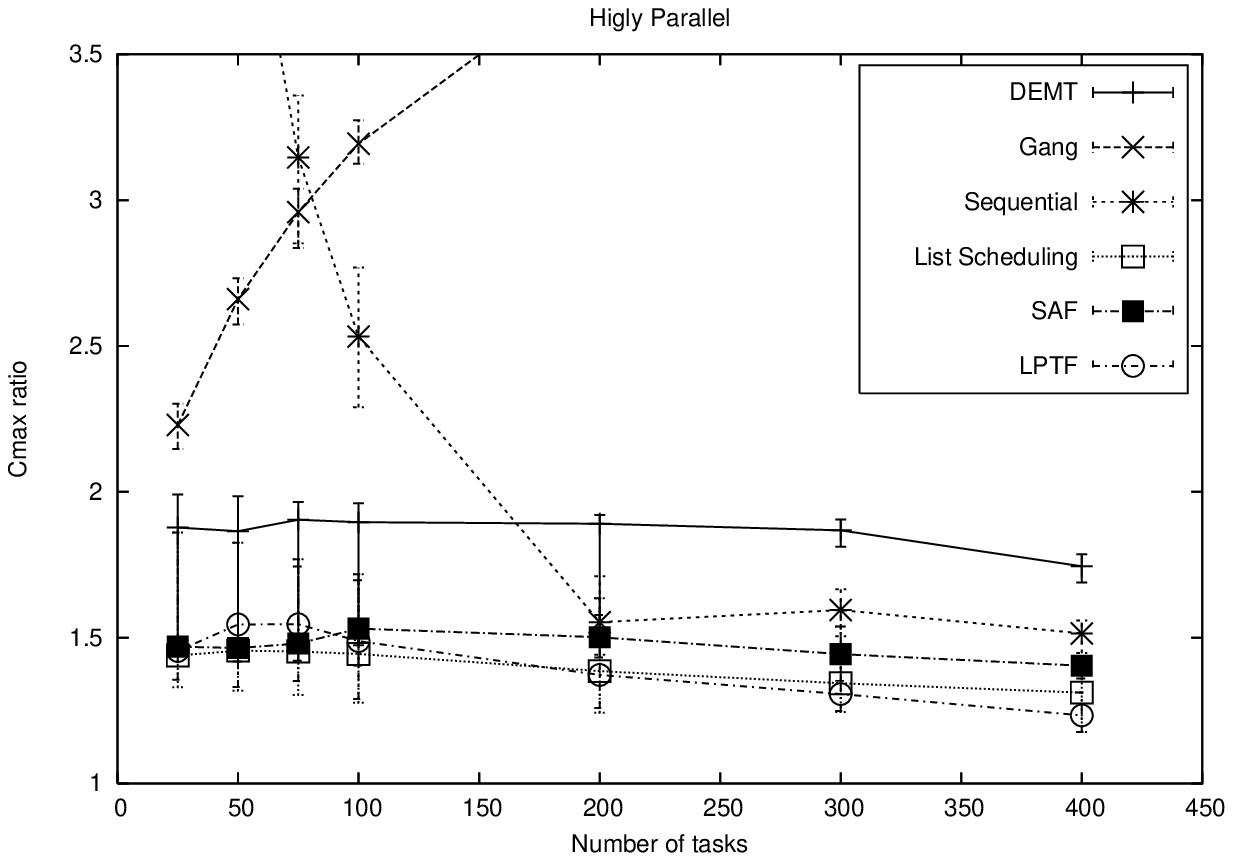}
\end{center}
\caption{Performance ratio for the simulation on 200 processors, highly parallel tasks}
\label{fig:courbe_hight}
\end{figure}

\begin{figure}[p]
\begin{center}
  \includegraphics[scale=0.70]{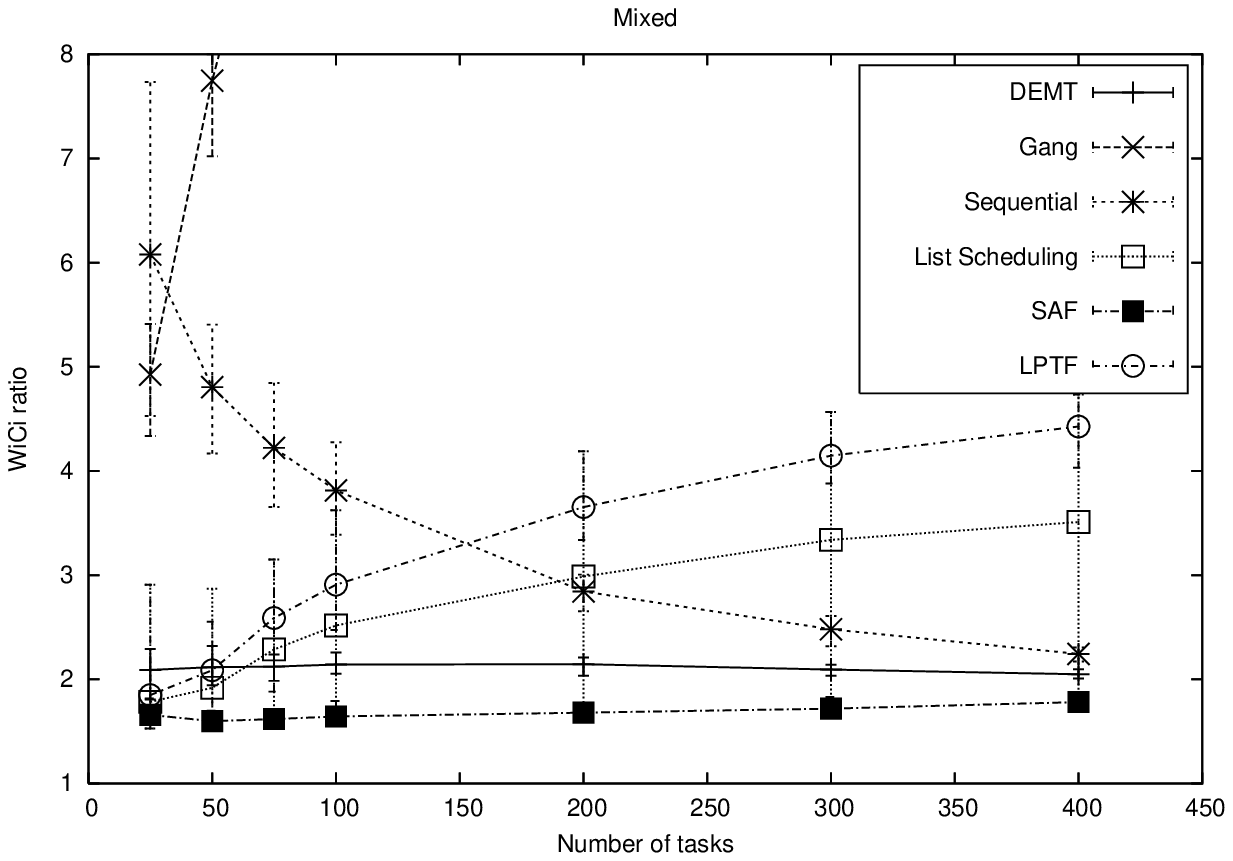}
  \includegraphics[scale=0.70]{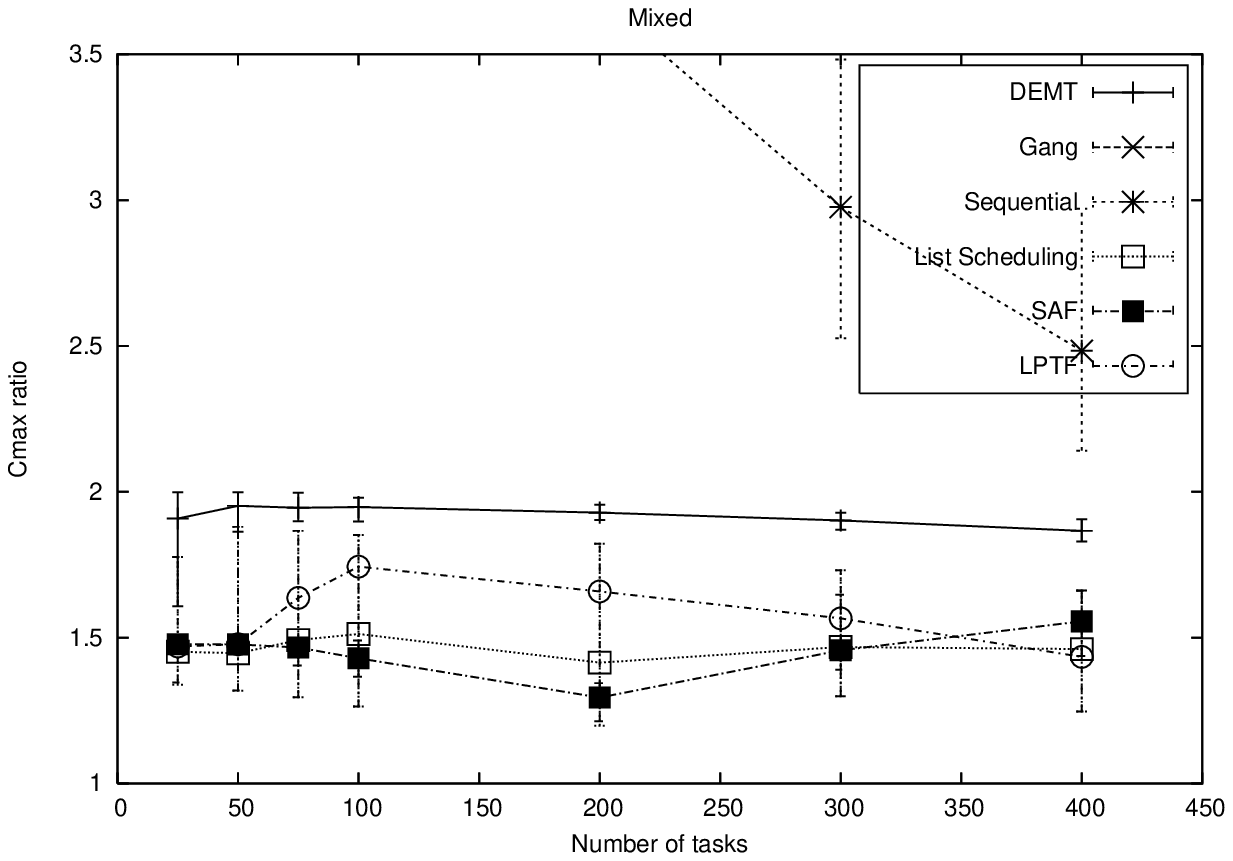}
\end{center}
\caption{Performance ratio for the simulation on 200 processors, mixed model parallel tasks}
\label{fig:courbe_mixed}
\end{figure}

Figure \ref{fig:courbe_hight} presents the same experiments with the
highly parallel tasks. On the minsum criterion, our algorithm is
clearly the best one. Gang and sequential have opposite behavior on
both criteria, Gang being good with a small number of tasks and
sequential good for a large number of tasks only. The other algorithms
are stable (with respect to the number of tasks) but with a larger
ratio on the minsum.
Remark that the allotment computed for list
algorithms is quite good, as $C_{max}$ performance ratio of these
algorithms is always smaller than $2$.

The next experiment (cf Figure \ref{fig:courbe_mixed}) presents mixed
instances with some large tasks and plenty of small tasks. In this
cases our algorithm is still quite stable with a performance ratio of
around $2$ for both criterion, however SAF is better than our
algorithm. The ratio of the two other list algorithms greatly increase
with the number of tasks, which points out that the order of tasks is
very important here.

Finally, the last experiment use a well known workload generator which
emulates real applications~\cite{CirneBerman}.  In this more realistic
setting our algorithm clearly outperforms the other ones for the
minsum criterion, and is also the only one to keep a stable ratio for
any number of tasks.

Several observations can be made from these results. First, the
performance ratio for the minsum criterion is never more than $2.5$,
and is on average around $2$. The performance ratio for the makespan
is almost always below $2$, and is $1.9$ on average. This is very
good, even for each criterion separately.

\begin{figure}[htbp]
\begin{center}
  \includegraphics[scale=0.70]{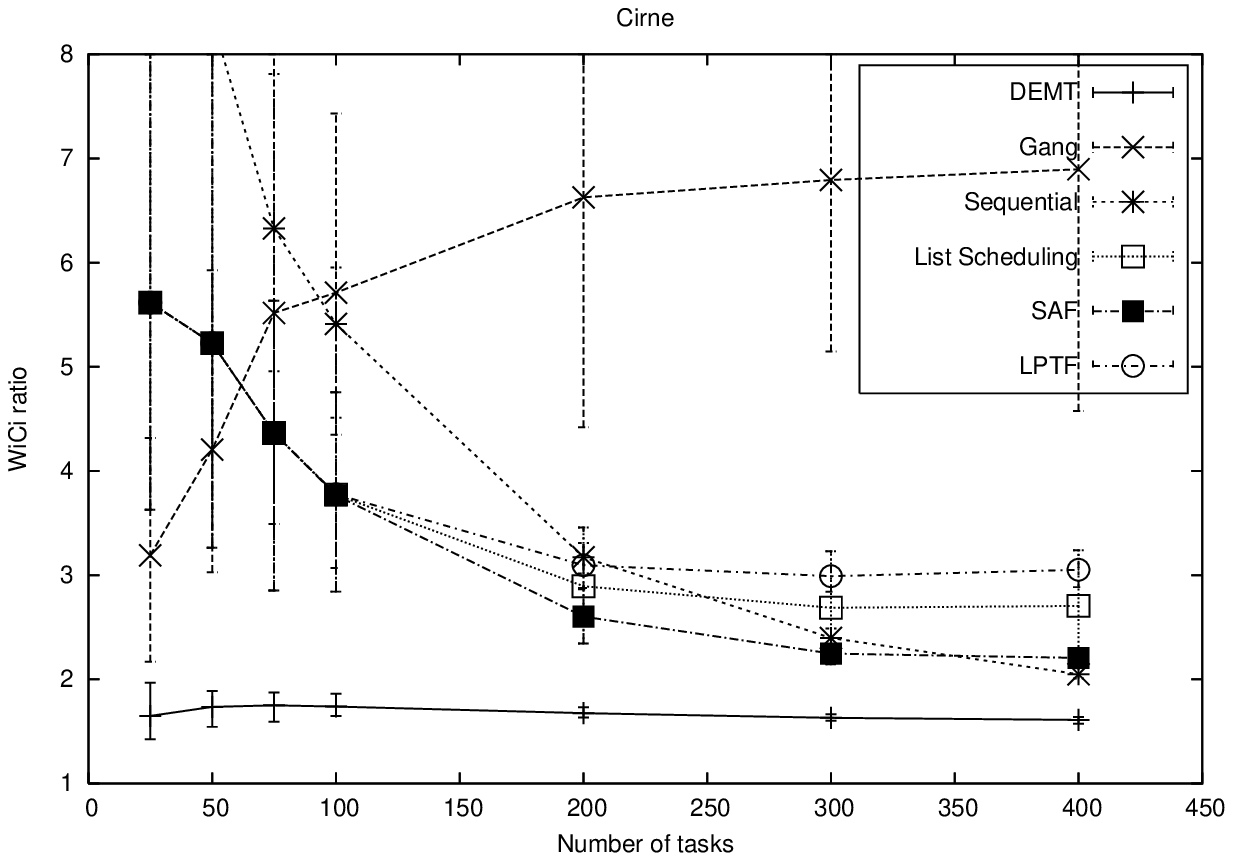}
  \includegraphics[scale=0.70]{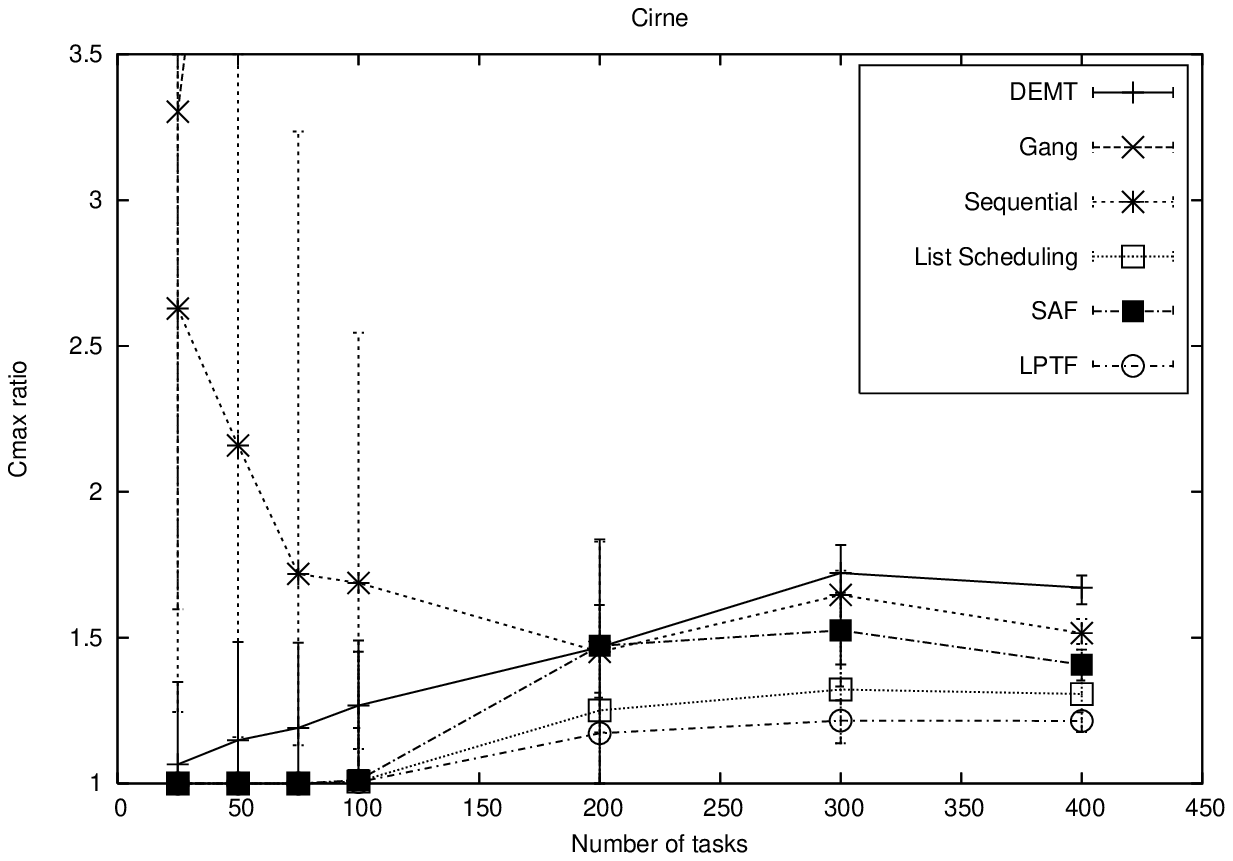}
\end{center}
\caption{Performance ratio for the simulation on 200 processors, cirne model parallel tasks}
\label{fig:courbe_cirne}
\end{figure}

The second observation is that our algorithm performs better when
tasks are more parallel. This can be understood if we remark that, for
a weakly parallel task, there is only one or two intervals in which it
can be scheduled without degrading its performance. So the scheduling
algorithm is more constrained when the tasks are not parallel.

The SAF algorithm perform quite well on simple cases. It appears on
complex cases that our approach is required to keep a good performance
on the minsum criterion. Thus our algorithms should be preferred in
actual applications as its performance ratio for minsum is insensitive
to jobs behavior and its performance ratio for the makespan is not far
from alternatives.

\begin{figure}[ht]
\begin{center}
\includegraphics[scale=0.70]{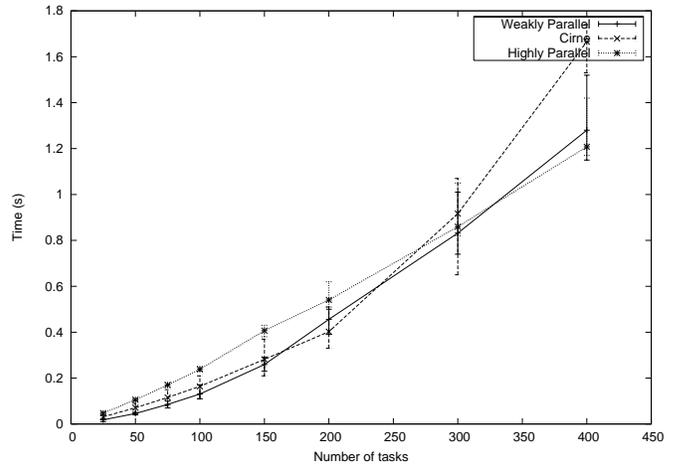}
\end{center}
\caption{Execution time of the algorithm.} \label{fig:time}
\end{figure}

Finally, Figure~\ref{fig:time} shows that the execution time of our
scheduling algorithm is low (less than 2 seconds for the largest
instances), as expected.

\section{Concluding remarks}
\label{sec:conclu}

In this paper we presented a new algorithm for scheduling a set of
independent jobs on a cluster. The main feature is to optimize two
criteria simultaneously. The experiments show that in average the
performance ratio is very good, and the algorithm is fast enough for
practical use. The algorithm has been assessed by comparing the minsum
performance to a new lower bound based on the relaxation of an ILP,
and comparing the makespan performance to the best known
approximation. Actual results are not available at the moment, but we
are currently implementing this algorithm on a full-scale platform
(Icluster2).

Several technical problems still have to be solved for an even more
efficient practical solution, namely the reservation of nodes which
reduces the size of the cluster and the mix of different types of jobs
(moldable jobs, rigid jobs, and divisible load jobs).







\end{document}